\documentclass[]{revtex4}
\usepackage{graphicx}
\usepackage{amsmath}
\usepackage{textcomp}
\usepackage{mathrsfs}
\usepackage{amsfonts}

\begin{document}

\newcommand{\vett}[1]{\mathbf{#1}}
\newcommand{\uvett}[1]{\hat{\vett{#1}}}
\newcommand{\beq}{\begin{equation}}
\newcommand{\eeq}{\end{equation}}
\newcommand{\bseq}{\begin{subequations}}
\newcommand{\eseq}{\end{subequations}}
\newcommand{\barr}{\begin{eqnarray}}
\newcommand{\earr}{\end{eqnarray}}
\newcommand{\GH}{Goos-H$\ddot{\mathrm{a}}$nchen }
\newcommand{\IF}{Imbert-Fedorov }
\newcommand{\bra}[1]{\langle #1|}
\newcommand{\ket}[1]{| #1\rangle}
\newcommand{\expectation}[3]{\langle #1|#2|#3\rangle}
\newcommand{\braket}[2]{\langle #1|#2\rangle}

\title{Goos-H\"anchen and Imbert-Fedorov Shifts for Epsilon-Near-Zero Materials}

\author{Arttu Nieminen$^1$, Andrea Marini$^2$, and Marco Ornigotti$^{1,*}$}
\affiliation{$^1$ Photonics Laboratory, Physics Unit, Tampere University, FI-33720 Tampere, Finland}
\affiliation{$^2$ Department of Physical and Chemical Sciences, University of L'Aquila, Via Vetoio, 67100 L'Aquila, Italy}
\email{marco.ornigotti@tuni.fi}

\begin{abstract}
We investigate the reflection of a Gaussian beam impinging upon the surface of an epsilon-near-zero (ENZ) medium. In particular, we discuss the occurrence of Goos-H\"anchen and Imbert-Fedorov shifts. Our calculations reveal that spatial shifts are significantly enhanced owing to the ENZ nature of the medium, and that their value and angular position can be tuned by tuning the plasma frequency of the medium. 
\end{abstract}

\maketitle

\section{Introduction}
Reflection of light from an interface is a very common and daily-life phenomenon, which can be very well understood using ray optics and the very well known Snell's law \cite{ref1}. However, when the wave properties of light are taken into account, like in the case, for example, of an optical beam impinging upon a reflective surface, deviations from Snell's law may appear. These are the celebrated Goos-H\"anchen (GH) \cite{ref2,ref3,ref4}, and Imbert-Fedorov (IF) \cite{ref5,ref6,ref7,ref8} shifts, occurring in the plane of incidence and in the plane perpendicular to it, respectively. GH and IF shifts have been studied for various beam configurations, including Hermite-Gaussian \cite{ref18}, Laguerre-Gaussian \cite{ref19,ref19a}, Bessel \cite{ref20}, and Airy \cite{ref21} beams, to name a few, and reflecting interfaces, such as dielectric \cite{ref22}, metallic \cite{ref23,ref24}, multilayered \cite{ref25,ref26}, and graphene coated \cite{ref27} surfaces, as well as in other contexts, such as optical waveguides \cite{ref28, ref29}, metamaterials \cite{ref29a}, and double barrier fermionic systems \cite{ref30}. Moreover,  the link between optical beam shifts and quantum mechanical weak measurements has also been pointed out \cite{ref31, ref32}, opening the way for novel and efficient way to measure such quantities in optics \cite{ref33,ref34}. A comprehensive review of the topic can be found in Ref. \cite{ref35}.

On the other hand, since the early 2000s,  metamaterials, i.e., artificially constructed media that show peculiar properties, such as negative refraction \cite{ref36}, have gained a lot of interest in the community for their potential of representing the basis for the next generation of photonic circuits \cite{ref37}. An interesting family of metamaterials are the so-called epsilon-near zero (ENZ) materials, characterised by the fact, that for a given spectral range of frequencies, the real part of their dielectric function vanishes, while the imaginary part remains finite \cite{ref38}. Typically, ENZ media can be either found in nature in the form of suitably doped semiconductors \cite{ref38a}, or artificially realized in terms of plasmonic structures at optical frequencies \cite{ref38b}. In the past years, ENZ materials have found interesting applicaitons in different areas of photonics, ranging from the realisation of efficient sub-wavelength imaging \cite{ref39, ref40}, directional emission \cite{ref41}, cloaking devices \cite{ref42}, and energy squeezing in narrow channels \cite{ref43}, to name a few. Moreover, spin-orbit interaction in transmission of vortex light beams through ENZ media has been also recently investigated \cite{ref44}.

Interestingly enough, despite the very intriguing properties that ENZ materials possess, no thorough investigation of the dynamics of GH and IF shifts for light beams reflecting upon ENZ surfaces has been carried out yet. The only exception to this, to the best of our knowledge, is provided in Ref. \cite{GH_ENZ}, where the spatial GH shift for light impinging upon an ENZ interface in total internal reflection regime has been numerically investigated. 

In this work, we present a complete theory of GH and IF shifts for a Gaussian beam impinging upon the surface of an ENZ medium. In particular, we show how the beam shifts are amplified by the ENZ character of the reflecting surface, provided that the wavelength of the impinging Gaussian beam is close enough to the plasma wavelength characteristic of the ENZ medium. 

This work is organised as follows: in Sect. 2, we review the usual formalism of the beam centroid, to calculate the GH and IF shifts, as a function of the reflection coefficients of the reflecting surface. Then, in Sect. 3, we specify the analysis to the case of a Gaussian beam impinging upon the surface of an ENZ material, which we describe within the limits of validity of the Drude model. In this section, moreover, we also discuss the salient features induced by the peculiar nature of ENZ materials, and, in particular, the possibility to use spatial GH shifts to sort different polarisation states of light efficiently. Finally, conclusions are drawn in Sect. 4.

\section{Beam Centroid and Beam Shifts}
In this section, we briefly review the beam centroid formalism, typically used to calculate the GH and IF shifts, for a beam of light reflecting from a surface. A detailed derivation of this formalism can be found, for example, in Ref. \cite{ref35}. 

\begin{figure}[t!]
\centering
\includegraphics[width=0.8\textwidth]{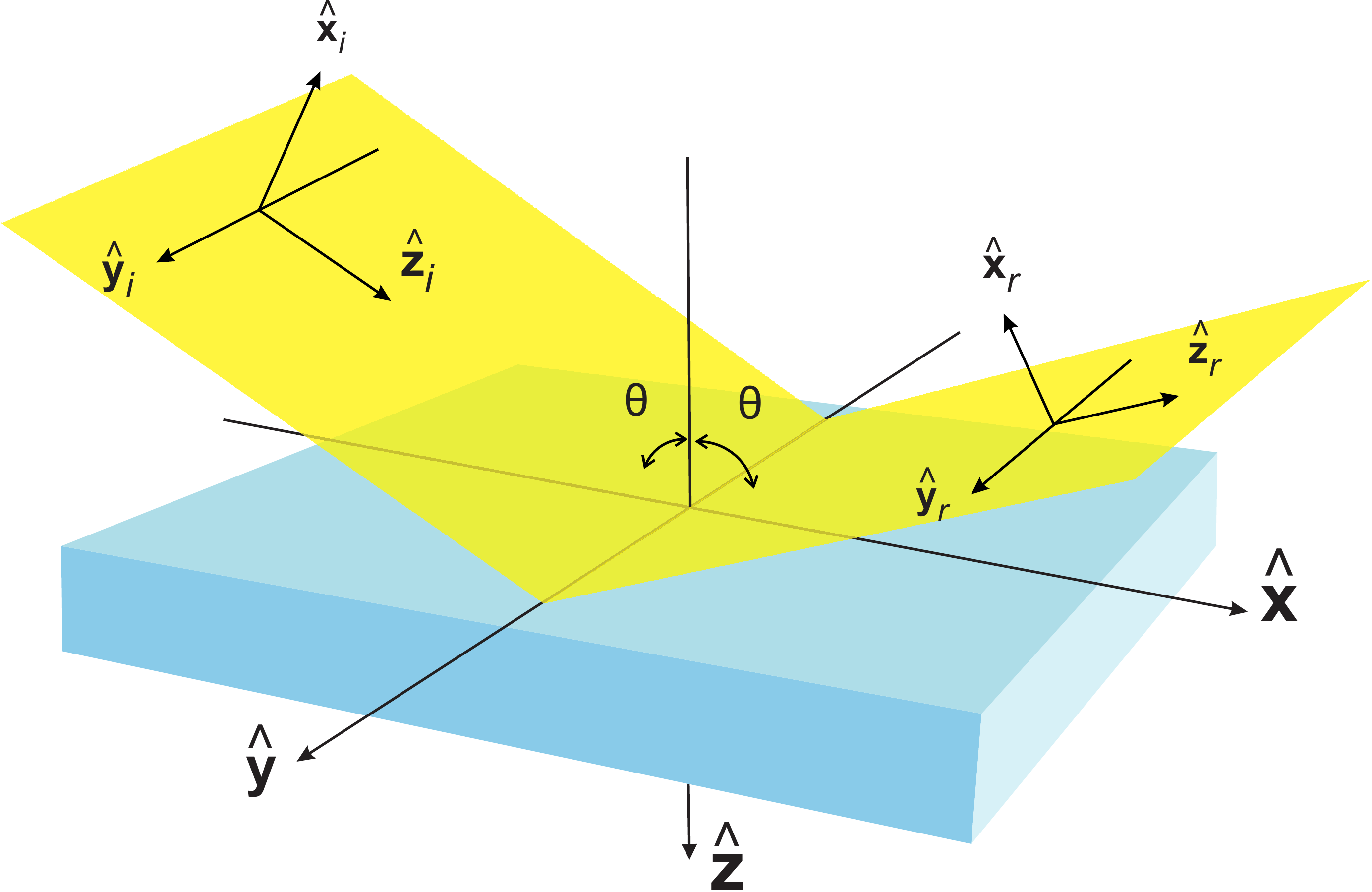}
\caption{\label{fig1} Schematic representation of the reflection geometry investigated in this paper.}
\end{figure}

To start our analysis, let us consider a monochromatic, paraxial electric field, impinging upon a surface, which, at this level of analysis, is only characterised by the usual Fresnel reflection coefficients $r_{\lambda}(\theta)$, for $\lambda=\{p,s\}$ polarisation \cite{ref1}. According to Fig. \ref{fig1}, we then define the laboratory frame $\{\uvett{x},\uvett{y},\uvett{z}\}$, and the incident, and reflected, frames $\{\uvett{x}_{i,r},\uvett{y}_{i,r},\uvett{z}_{i,r}\}$, atteched, respectively, to the incident, and reflected, beam. The impinging electric field can be then written as follows
\beq
\vett{E}_I(\xi,\eta,\zeta)=\sum_{\lambda=1}^2\,\int\,d^2K\,\uvett{u}_{\lambda} (U,V;\theta)\alpha_{\lambda}(U,V;\theta)\mathcal{E}(U,V)e^{iW\zeta}e^{i(U\xi+V\eta)},
\eeq

where $\{\xi,\eta,\zeta\}$ are the normalised cartesian coordinates,  $\lambda=\{1,2\}$ is a polarisation index (corresponding to $p$, and $s$, polarisation, respectively), $\{U,V,W=\sqrt{1-U^2-V^2}\}$ are the dimensionless components of the $\vett{k}$-vector in the incident frame, $\uvett{u}_{\lambda}$ are the local polarisation vectors, attached to the single plane wave components of the beam, $\alpha_{\lambda}=\uvett{f}\cdot\uvett{u}_{\lambda}$ are the polarisation functions (with $\uvett{f}=a_p\uvett{x}+a_Se^{i\Delta}\uvett{y}$, and $a_P,a_S,\Delta\in\mathbb{R}$, with the constraint $a_P^2+a_S^2=1$), and $\mathcal{E}(U,V)$ is the beam spectrum. Upon reflection, the electric field in the reflected frame can be obtained by simply applying the law of specular reflection on the local basis. This corresponds to set $\uvett{u}_{\lambda}(U,V,W;\theta)\rightarrow\uvett{u}_{\lambda}(-U,V,W;\pi-\theta)$ and $\mathcal{E}(U,V)\rightarrow r_{\lambda}(U,V)\mathcal{E}(U,V)$, where $r_{\lambda}(U,V)$ are the reflection coefficients of the surface. In the reflected frame, then, the electric field can be written as follows: 
\beq
\vett{E}_R(\xi,\eta,\zeta)=\sum_{\lambda=1}^2\,\int\,d^2K\,\uvett{u}_{\lambda} (-U,V;\pi-\theta)\alpha_{\lambda}(U,V;\theta)r_{\lambda}(U,V)\mathcal{E}(U,V)e^{iW\zeta_r}e^{i(-U\xi_r+V\eta_r)}.
\eeq

Beam shifts can be then evaluated by computing the beam centroid in the reflected frame, i.e., $\langle\vett{X}\rangle=\langle\xi_r\rangle\uvett{x}+\langle\eta_r\rangle\uvett{y}$, where, using a quantum-like formalism introduced for beam shifts \cite{ref35}
\beq
\langle \xi_r\rangle =\frac{\langle  \vett{E}_R|\xi_r| \vett{E}_R\rangle}{\langle\vett{E}_R| \vett{E}_R\rangle},\hspace{1cm}\langle \eta_r\rangle =\frac{\langle  \vett{E}_R|\eta_r| \vett{E}_R\rangle}{\langle  \vett{E}_R| \vett{E}_R\rangle}.
\eeq
 The spatial ($\Delta$) and angular ($\Theta$) shifts can be then written in the following form \cite{ref35}:
\bseq
\begin{align}
\Delta_{GH}=\langle \xi_r\rangle|_{\zeta=0}, \hspace{1cm} \Theta_{GH}=\frac{\partial\langle \xi_r\rangle}{\partial\zeta}\Big|_{\zeta=0},\\
\Delta_{IF}=\langle \eta_r\rangle|_{\zeta=0}, \hspace{1cm} \Theta_{IF}=\frac{\partial\langle \eta_r\rangle}{\partial\zeta}\Big|_{\zeta=0}.
\end{align}
\eseq
For the case of a Gaussian beam impinging onto a reflecting surface, characterised by standard Fresnel coefficients $r_{\lambda}(\theta)$, the formulas above give the following result, which will be useful for our analysis in the next section
\bseq\label{shifts1}
\begin{align}
k_0\Delta_{GH}&=\sum_{\lambda=1}^2w_{\lambda}\frac{\partial\phi_{\lambda}}{\partial\theta}\label{spGHeq},\\
\zeta_Rk_0\Theta_{GH}&=\sum_{\lambda=1}^{2}w_{\lambda}\frac{\partial\ln R_{\lambda}}{\partial\theta},\label{anGHeq}\\
k_0\Delta_{IF}&=-\Big\{\cot\theta\Big[\frac{w_pa_s^2+w_sa_p^2}{a_pa_s}\cos\Delta\nonumber\\
&+ 2\sqrt{w_pw_s}\sin(\Delta+\phi_s-\phi_p)\Big]\Big\},\\
\zeta_Rk_0\Theta_{IF}&=\left[\frac{w_pa_s^2-w_sa_p^2}{a_pa_s}\cos\Delta\cot\theta\right],
\end{align}
\eseq
where $k_0$ is the vacuum wave vector, $\zeta_R$ is the Gaussian beam Rayleigh range, $w_{\lambda}=a_{\lambda}^2R_{\lambda}^2/(a_p^2R_p^2+a_sR_s^2)$ is the fraction of energy contained in each polarisation, and $r_{\lambda}=R_{\lambda}e^{i\phi_{\lambda}}$ has been assumed.

\section{Goos-H\"anchen and Imbert-Fedorov Shifts for ENZ Media}
\begin{figure}[t!]
\centering
\includegraphics[width=0.8\textwidth]{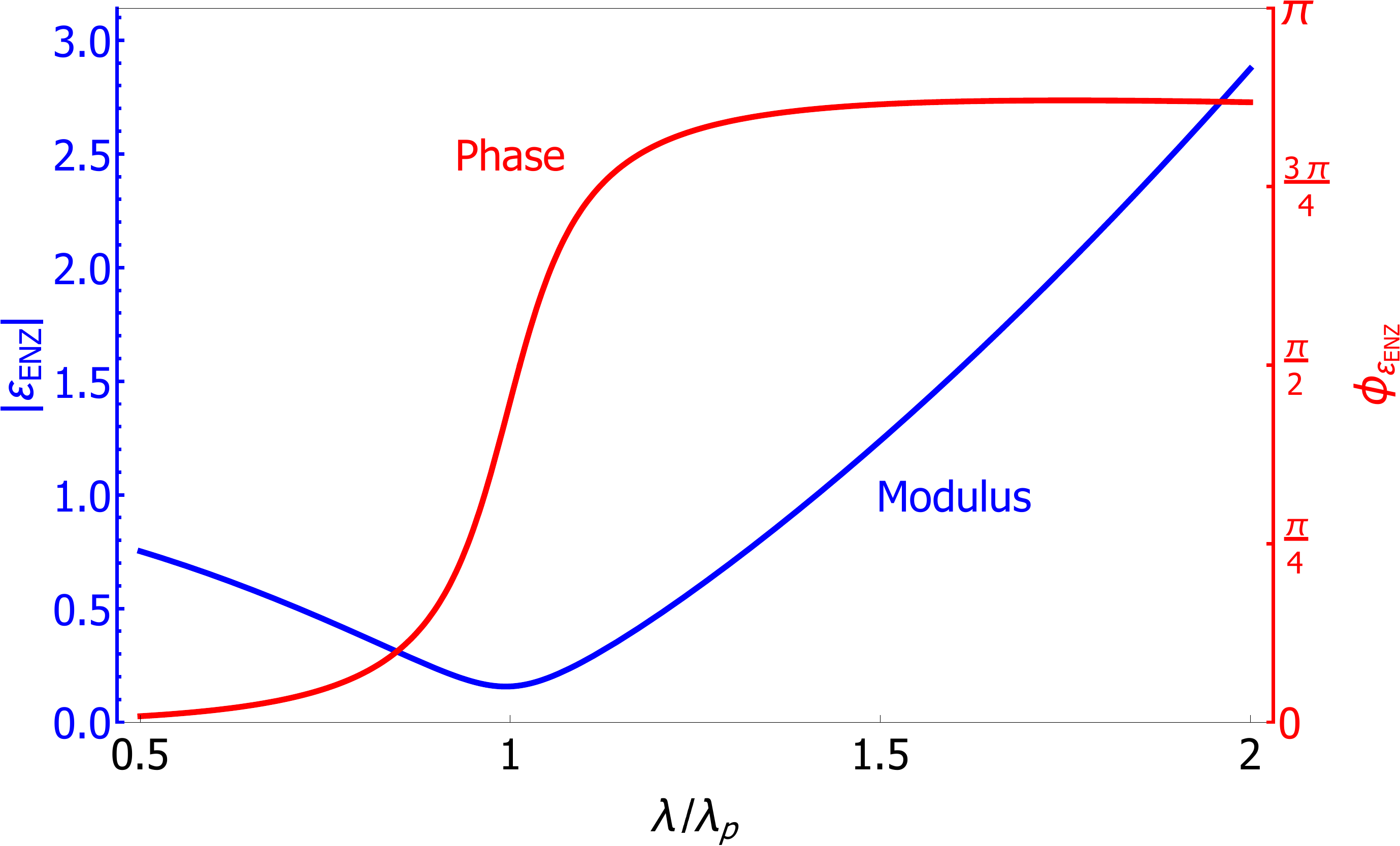}
\caption{\label{figure2} Modulus (blue, solid, line), and phase (red, solid, line) of the dielectric function $\varepsilon_{ENZ}(\lambda)$, for an ENZ medium. As it can be seen, the modulus of the dielectric function reaches its minimum value, for $\lambda\simeq\lambda_p$. Notice, moreover, that in the range of wavelengths corresponding to $0.5<\lambda/\lambda_p<1.2$, we have $|\varepsilon_{ENZ}(\lambda)|<1$. This corresponds, to the ENZ region of this material. In this interval of wavelength, therefore, we expect ENZ effects to be prominent. For this figure, a plasma wavelength of $\lambda_p=3$ $\mu m$, and a damping coefficient of $\Gamma= 5\times 10^4$ $m^{-1}$, corresponding to $\gamma=10^{14}$ Hz.}
\end{figure}
To estimate GH and IF for Gaussian beams impinging upon ENZ materials, we first need to take a closer look to the reflection coefficients (for both p- and s-polarisation) of an ENZ material. To do that, we first need to find an explicit expression for the dielectric function of an ENZ medium, as a function of the wavelength of the impinging light. Within the Drude approximation \cite{DrudeENZ}, this is given by
\beq\label{epsilonENZ}
\varepsilon_{ENZ}(\lambda)=1-\frac{1}{\lambda_p^2}\left[\frac{1}{\lambda}\left(\frac{1}{\lambda}+i\Gamma\right)\right]^{-1},
\eeq
where  $\lambda_p=2\pi c/\omega_p$ is the plasma wavelength of the medium, and $\Gamma=\gamma/2\pi c$ is the spatial damping parameter, related to the Drude damping factor $\gamma$, typically given in terms of fractions of the plasma frequency $\omega_p$. The dependence of the modulus and phase of the dielectric function on the wavelength is shown in Fig. \ref{figure2}. As it can be seen, as $\lambda$ approaches the plasma wavelength $\lambda_p$, characteristic of the particular ENZ medium, the modulus of the dielectric function $|\varepsilon_{ENZ}(\lambda)|$ reaches its minimum value, and, moreover, $|\varepsilon_{ENZ}(\lambda_p)|\ll 1$. This is a signature, of the ENZ nature of the material considered in our analysis. In addition to that, it is worth noticing that in the range of wavelengths $0.5<\lambda/\lambda_p<1.2$ the modulus of $\varepsilon_{ENZ}$ remains smaller than one. Within this range of values, then, we expect ENZ effects to be prominent, and give a significant contribution to the beam shifts. Indeed, ENZ media are known to possess several peculiar properties ensuing from their large effective wavelength that leads to a marked slow-light regime \cite{Ale2013,Stockmann}. In particular, such a slow-light regime is responsible for enhanced spin-orbit interaction of light in ENZ media \cite{Ale2017}. Intuitively, such an effect can be understood through the inherently large effective wavelength inside ENZ media, which thus implies that externally impinging paraxial fields behave as highly non-paraxial fields inside ENZ materials, thus leading to a geometric enhancement of spin-orbit interaction of light \cite{Ale2017}.  

The reflection coefficients for both p- and s-polarisation can be then written, using Eq. \eqref{epsilonENZ}, in the traditional Fresnel form as follows:
\bseq
\begin{align}
r_p(\theta,\lambda)&=-\frac{\varepsilon_{ENZ}(\lambda)\cos\theta-\sqrt{\varepsilon_{ENZ}(\lambda)-\sin^2\theta}}{\varepsilon_{ENZ}(\lambda)\cos\theta+\sqrt{\varepsilon_{ENZ}(\lambda)-\sin^2\theta}},\\
r_s(\theta,\lambda)&=\frac{\cos\theta-\sqrt{\varepsilon_{ENZ}(\lambda)-\sin^2\theta}}{\cos\theta+\sqrt{\varepsilon_{ENZ}(\lambda)-\sin^2\theta}}.
\end{align}
\eseq
\subsection{Spatial GH Shift}
\begin{figure}[t!]
\includegraphics[width=1.0\textwidth]{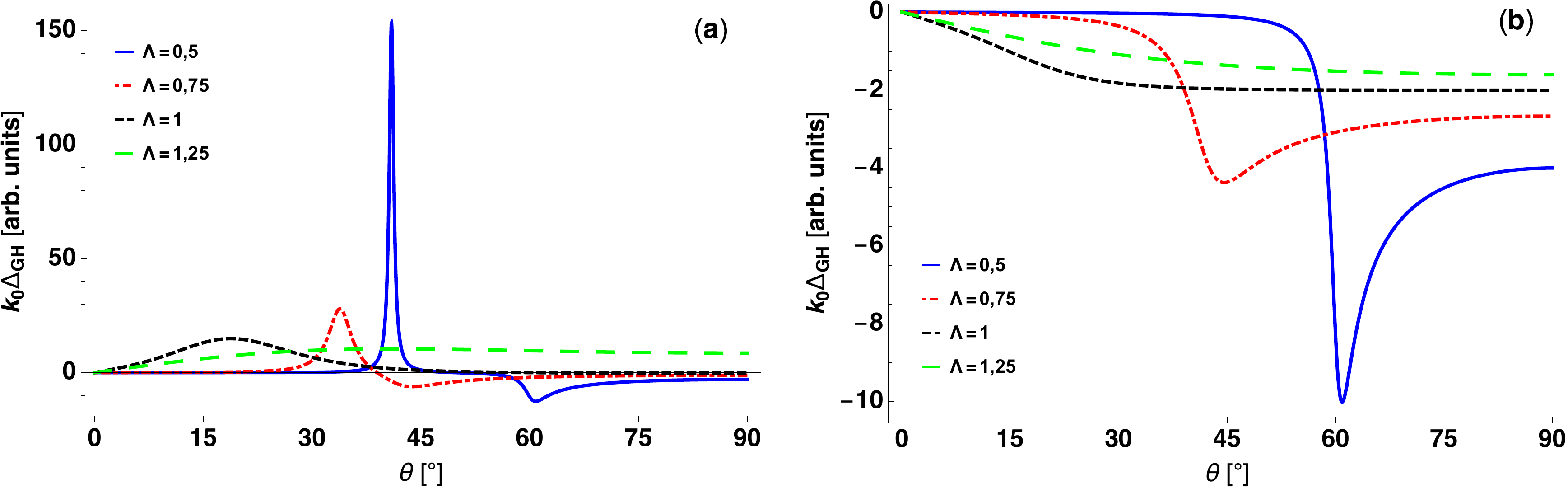}
\caption{Normalised spatial GH shifts for (a) p-polarisation, and (b) s-polarisation, for different values of the normalised wavelength $\Lambda=\lambda/\lambda_p$. For both polarisations, the spatial shifts are nonzero, since the reflection coefficients of ENZ media are intrinsically complex, and therefore $\partial\phi_{p,s}/\partial\theta\neq 0$, even away from total internal reflection. Notice, in particular, that while $\Delta_{GH}$ for s-polarisation is always negative [panel (b)], for p-polarisation it can be either positive or negative, for $\Lambda<1$, while it is always positive, for $\Lambda\geq 1$.  Notice, that the magnitude of the black (dot-dashed), and green (dashed) curves in panel (a) has been artificially enhanced by a factor of 10, in order to allow to visualise the plots for various values of $\Lambda$ in a single figure, in a neat, and clear way.}
\label{figure3}
\end{figure}

The (normalised) spatial GH shift $k_0\Delta_{GH}$ for $p$- polarisation [panel (a)], and $s-$polarisation [panel (b)], as a function of the normalised wavelength  $\Lambda=\lambda/\lambda_p$, of the impinging Gaussian beam, onto the ENZ surface, is depicted in Fig. \ref{figure3}. First of all, notice that for both polarisations, $k_0\Delta_{GH}\neq 0$, despite the reflection is occurring from air-to-ENZ medium. In a standard setting, with reflection occurring, say, at an air-to-glass interface, the reflection coefficient is a purely real number, and the spatial GH shift is zero, according to Eq. \eqref{spGHeq}, since $\phi_p=\phi_s=0$. In this case, however, the reflection coefficient is a complex quantity, and therefore, even for air-to-ENZ reflection, $\phi_{p,s}(\theta,\lambda)\neq 0$, and a spatial GH shift can occur.
For the case of $p$-polarisation, Fig. \ref{figure3}(a) reveals that the spatial shift oscillates between positive and negative values, depending on the incidence angle and the wavelength. For $s$-polarisation, on the other hand, we have the opposite situation, where the shift is always negative, as it can be seen from Fig. \ref{figure3}(b). Moreover, we observe a clear resonance structure of $k_0\Delta_{GH}$ for both polarisations. This resonance occurs, in particular, close to Brewster angle, for $p$-polarisation, and it reaches its maximum value for $\Lambda=0.5$, i.e., for an impinging Gaussian beam with a wavelength equal to half the plasma wavelength of the ENZ medium. For the case of Fig. \ref{figure3}(a), $\lambda_p=3$ $\mu m$ has been used. In these conditions, an impinging Gaussian beam at wavelength $\lambda=1.5$ $\mu$m (solid blue line), the maximum spatial GH shift is achieved around  $\theta_{inc}\simeq 45^{\circ}$ and it corresponds to $\Delta_{GH}=150/k_0\simeq23\lambda\simeq 35$ $\mu m$. For a Gaussian beam at wavelength $\lambda=2.25$ $\mu$m (red, dashed line in Fig. \ref{figure3}(a), corresponding to $\Lambda=0.75$), on the other hand, the maximum spatial GH shift occurs at smaller incidence angles, i.e., around $\theta_{inc}\simeq 33^{\circ}$, and its magnitude is also reduced to $\Delta_{GH}=30/k_0\simeq 4.8\lambda\simeq 11$ $\mu m$. 

For the case of $s-$polarisation, for $\Lambda=0.5$ the resonance peak [solid, blue line in Fig. \ref{figure3}(b)] occurs at greater incidence angles, i.e., around $\theta_{inc}\simeq 60^{\circ}$, and it corresponds to a spatial GH shift of $\Delta_{GH}=10/k_0\simeq 1.6\lambda\simeq 2.4$ $\mu$m, while for $\Lambda=0.75$  [dashed, red line in Fig. \ref{figure3}(b)] it occurs around $\theta_{inc}\simeq 40^{\circ}$ with a magnitude of $\Delta_{GH}=5/k_0\simeq 0.8\lambda=1.8$ $\mu$m.

A closer look to Figs. \ref{figure3}(a) and (b) reveals another interesting detail. Indeed, while at Brewster incidence (and for an impinging beam characterised by $\lambda=0.5\lambda_p$), in fact, the spatial GH shift for $p-$ polarisation undergoes a maximum value of $\simeq150/k_0$, at the same angle of incidence $\Delta_{GH}\simeq 0$ for $s-$polarisation. This result is quite interesting, as it suggests that spatial GH shifts from ENZ surfaces can be used as an efficient method to spatially sort the polarisation states of the impinging light. The significant shift of about 25 wavelegths experienced by $p-$polarisation at Brewster incidence, compared to the negligible one of $s-$ polarisation, in fact, can become large enough to spatially resolve the two polarisation states upon reflection. In this respect, the electrical tunability of the plasma frequency of ENZ materials \cite{Howard} could provide an active mean of controlling the separation between spots of distinct polarization states, thus providing a viable platform towards the development of integrated polarisation sorters. 
\begin{figure}[t!]
\includegraphics[width=1.0\textwidth]{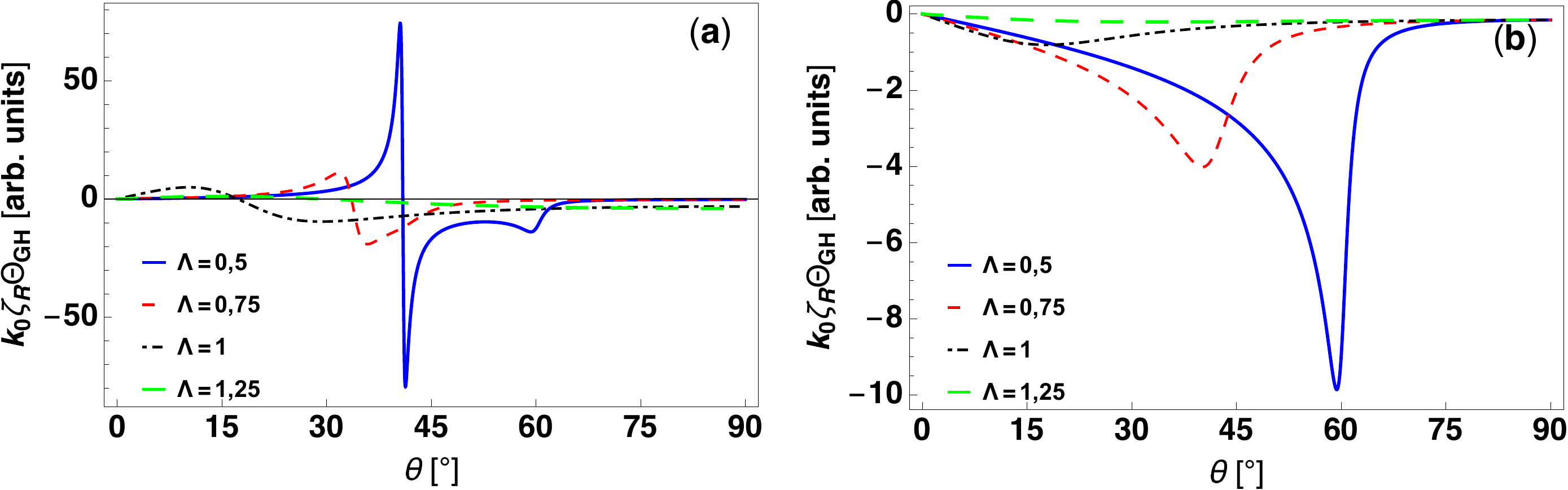}
\caption{Normalised angular GH shift for (a) p-polarisation, and (b) s-polarisaiton, for different values of the normalised wavelength $\Lambda=\lambda/\lambda_p$. Notice, that while for p-polarisation, the value of $\Theta_{GH}$ oscillates between positive, and negative values, across the near-zero wavelength ($\Lambda=1$), for s-polarisation, instead, it is always negative, and approaches zero, as $\Lambda$ is increased beyond $\Lambda=1$. Also, the maximum angular shift for s-polarisation [panel (b)] drastically shifts towards lower incidence angles, as $\Lambda\rightarrow 1$. Notice, that the magnitude of the black (dot-dashed), and green (dashed) curves in panel (a) has been artificially enhanced by a factor of 10, in order to allow to visualise the plots for various values of $\Lambda$ in a single figure, in a neat, and clear way.}
\label{figure4}
\end{figure}
\subsection{Angular GH Shift}
Th angular GH shift is depicted in Fig. \ref{figure4} for $p-$ [panel (a)], and $s-$polarisation [panel (b)]. It is interesting to notice, that despite the expression of the angular shift in Eq. \eqref{anGHeq} has been derived by using the standard, first order expansion of the reflection coefficients, with respect to $U$, and $V$ \cite{ref32}, its value at Brewster's angle does not diverge, as expected, but remains finite. This is due to the fact, that while $\operatorname{Re}\{r_p(\theta_b,\lambda)\}=0$ at the Brewster angle $\theta_b$, \hspace{0.5mm} $\operatorname{Im}\{r_p(\theta_b,\lambda)\}\neq 0$ (although very small), which, in turn, implies that $|r_p(\theta_b,\lambda)|\neq 0$. This, ultimately, acts as a renormalisation of the logarithmic derivative appearing in Eq.  \eqref{anGHeq} , which is now regular, and does not diverge anymore at the Brewster angle. Moreover, as the wavelength of the impinging Gaussian beam becomes larger than the plasma wavelength of the ENZ medium (namely, as $\Lambda$ becomes larger than 1),  the angular shifts for both polarisation tend to disappear [see, for exmaple, the dashed, green curve, in both panels of Fig. \ref{figure4}]. This is due to the fact, that for $\Lambda\gg1$, $\varepsilon_{ENZ}\simeq 1-\Lambda^2$, and the modulus of the reflection coefficient is approximately $R_{p,s}\simeq 1$. According to Eq. \eqref{anGHeq}, the angular GH shift is related to the derivative of $R_{p,s}$ with respect to $\theta$, which in this limit approaches zero, because the modulus of the reflection coefficient is constant with respect to $\theta$.
\subsection{IF Shifts}
\begin{figure}[t!]
\centering
\includegraphics[width=\textwidth]{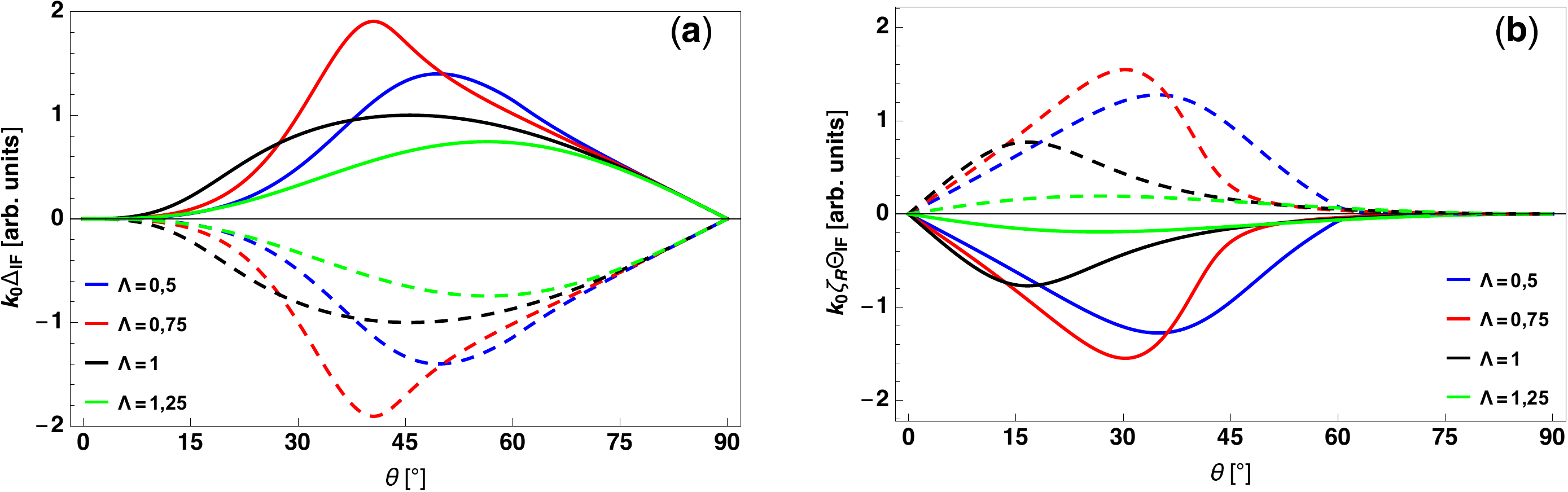}
\caption{Normalised spatial (a), and angular (b) IF shift for different values of the normalised wavelength $\Lambda=\lambda/\lambda_p$. The solid (dashed) lines in panel (a) represent left (right)-handed circular polarisation, while in panel (b) they represent diagonal (anti-diagonal) polarisation. Notice, moreover, how $\Theta_{IF}$ [panel (b)] vanishes, for incidence angles $\theta>60^{\circ}$. This is in contrast to standard results, where $\Theta_{IF}=0$, only in the very vicinity of normal incidence.}
\label{figure5}
\end{figure}

The spatial (a), and angular (b) IF shifts, are depicted in Fig. \ref{figure5}, for circular, and diagonal, polarisation, respectively. In agreement with the standard theory of IF shift, the spatial shift is nonzero only for circularly polarised light. Moreover, the shifts for left-handed [solid lines in Fig. \ref{figure5}(a)], and right-handed [dashed lines in Fig. \ref{figure5}(b)] circular polarisation are mirror images of each other. No significant differences in the structure of the spatial IF shifts for ENZ, compared to standard dielectrics, however, is to be noticed. For the angular IF shift, on the other hand, despite the fact its global structure reflects the traditional one, that is observed from dielectric surfaces, it is interesting to note, that $\Theta_{IF}=0$ for incidence angles $\theta>60^{\circ}$. This is in contrast, with the standard result, where for diagonal [solid line, in Fig. \ref{figure5}(b)], and anti-diagonal [dashed lines, in Fig. \ref{figure5}(b)], $\Theta_{IF}=0$ only in the very vicinity of $\theta=90^{\circ}$.

\section{Conclusions}
In this work, we have presented a comprehensive analysis of GH, and IF shifts for Gaussian beams impinging upon ENZ media, and we have discussed how the ENZ character affects the beam shifts. In particular, for spatial GH shifts, we have predicted a significant magnification of $\Delta_{GH}$, which reaches its maximum of about $25\lambda$ for $p-$polarisation, and about $2\lambda$ for $s-$polarisation, for a Gaussian beam characterised by a wavelength $\lambda=0.5\lambda_p$. Moreover, at Brewster incidence, $\Delta_{GH}^{(p-polarisation)}\gg\Delta_{GH}^{s-polarisation}$. This suggests the possibility to use beam shifts at ENZ surfaces as a viable mean to control the separation between spots of distinct polarisation states. To make our analysis complete and self-consistent, we have also reported the spatial, and angular IF shifts, although their properties are not affected by ENZ, as much as it happens for the GH shift. Our results are then encouraging for the development of devices based on ENZ materials enabling efficient electrical control of radiation polarisation at the nanoscale.

\section*{Acknowledgements}
The work is part of the Academy of Finland Flagship Programme, Photonics Research and Innovation (PREIN), decision 320165. A. M. and M. O. thank G. Enziana for fruitful discussions.

\section*{References}


\begin{thebibliography}{99}
%
\bibitem{ref1} M. Born and E. Wolf, \emph{Principles of Optics} 7th edn (New York: Cambridge University Press)
%
\bibitem{ref2} Goos F and H\"anchen H 1947 \emph{Ann. Phys.} \textbf{1}, 333
%
\bibitem{ref3} K. Artmann 1948 \emph{Ann. Phys.} \textbf{2}, 87 
%
\bibitem{ref4} M. McGuirk and C. K. Carniglia 1977 \emph{J. Opt. Soc. Am.} \textbf{67}, 103 
%
\bibitem{ref5} F. I. Fedorov 1955 \emph{Dokl. Akad. Nauk SSSR} \textbf{105}, 465 
%
\bibitem{ref6} C. Imbert 1972  \emph{Phys. Rev. D} \textbf{5}, 787 
%
\bibitem{ref7} F. Pillon, H. Gilles and S. Girard 2004 \emph{Appl. Opt.} \textbf{43}, 1863 
%
\bibitem{ref8} H. Schilling 1965 \emph{Ann. Phys.} (Berlin) \textbf{16}, 122 
%
\bibitem{ref18} D. Golla and S. Dutta Gupta, \emph{arXiv:1011.3968v1}.
%
\bibitem{ref19} K. Y. Bliokh, I. V. Shadrivov and Y. S. Kivshar 2009 \emph{Opt. Lett.} \textbf{34}, 389 
%
\bibitem{ref19a} M. Merano, N. Hermosa, J. P. Woerdman and A. Aiello 2010 \emph{Phys. Rev. A} \textbf{82}, 023817 
%
\bibitem{ref20} A. Aiello and J. P. Woerdman 2011 \emph{Opt. Lett.} \textbf{36}, 543 
%
\bibitem{ref21} M. Ornigotti 2018 \emph{Opt. Lett.} \textbf{43}, 1411 
%
\bibitem{ref22} S. Kozaki and H. Sakurai 1978 \emph{J. Opt. Soc. Am.} \textbf{68}, 508
%
\bibitem{ref23} P. T. Leung, C. W. Chen and H. P. Chiang 2007 \emph{Opt. Commun.} \textbf{276}, 206
%
\bibitem{ref24} M. Merano, A. Aiello,  G. W. 't Hooft,  M. P. van Exter, E. R. Elier and J. P. Woerdman 2007 \emph{Opt. Expr.} \textbf{15}, 15928 
%
\bibitem{ref25} T. Tamir 1986 \emph{J. Opt. Soc. Am. A} \textbf{3}, 558 
%
\bibitem{ref26} G. D. Landry and T. A. Maldonado 1996 \emph{Appl. Opt.} \textbf{35}, 5870 
%
\bibitem{ref27} S. Grosche, M. Ornigotti and A. Szameit 2015 \emph{Opt. Express} \textbf{23}, 30195 
%
\bibitem{ref28} S. Grosche, M. Ornigotti and A. Szameit 2016 \emph{Phys. Rev. A} \textbf{94}, 063831 
%
\bibitem{ref29} A. Szameit, H. Trompeter, M. Heinrich, F. Dreisow, U. Peschel, T. Pertsch, S. Nolte, F. Lederer and A. T\"nnermann 2008 \emph{New J. Phys.} \textbf{10}, 103020 
%
\bibitem{ref29a}I. V. Shadrivov, A. A. Zharov and Y. S. Kivshar 2003 \emph{Appl. Phys. Lett.} \textbf{83}, 2713 
%
\bibitem{ref30}A. Jellala, I. Redouanic, Y. Zahidic and H. Bahloulia 2014 \emph{Physica E} \textbf{58}, 30 
%
\bibitem{ref31} M. R. Dennis and J. B. G\"otte 2012 \emph{New J. Phys.} \textbf{14}, 073013 
%
\bibitem{ref32} F. T\"oppel, M. Ornigotti and A. Aiello 2013 \emph{New J. Phys.} \textbf{15}, 113059 
%
\bibitem{ref33} G. Jayaswal, G. Mistura and M. Merano 2013 \emph{Opt. Lett.} \textbf{38}, 1232 
%
\bibitem{ref34} G. Jayaswal, G. Mistura and M. Merano 2014 \emph{Opt. Lett.} \textbf{39}, 2266 
%
\bibitem{ref35}  K. Y. Bliokh and A. Aiello 2013 \emph{J. Opt.} \textbf{15}, 014001 
%
\bibitem{ref36} Metamaterial Electromagnetic Cloak at Microwave Frequencies
D. Schurig1, J. J. Mock1, B. J. Justice1, S. A. Cummer1, J. B. Pendry2, A. F. Starr3, D. R. Smith1,* Science  10 Nov 2006: Vol. 314, Issue 5801, pp. 977-980.
%
\bibitem{ref37} Plasmonics: merging photonics and electronics at nanoscale dimensions, Ekmel Ozbay Science Vol. 311 Pages 189-193 (2006)
%
\bibitem{ref38} Near-zero refractive index photonics, Authors Inigo Liberal, Nader Engheta, Nature Photonics Volume 11 Pages 149 (2017)
%
\bibitem{ref38a} Alternative Plasmonic Materials: Beyond Gold and Silver Gururaj V. Naik  Vladimir M. Shalaev  Alexandra Boltasseva First published: 15 May 2013
%
%
\bibitem{ref38b} Wes, P. R. et al. 2010 Searching for better plasmonic materials \emph{Laser Photon. Rev.} \textbf{4}, 6 
%
\bibitem{ref39} A. Alu, M. G. Silveirinha, A. Salandrino and N. Engheta 2007 \emph{Phys. Rev. B} \textbf{75}, 155410 
%
\bibitem{ref40} G. Castaldi, S. Savoia, V. Galdi, A. Alu and N. Engheta 2012 \emph{Phys. Rev. B} \textbf{86}, 115123 
%
\bibitem{ref41} S. Enoch, G. Tayeb, P. Sabouroux, N. Guerin and P. Vincent 2002 \emph{Phys. Rev. Lett.} \textbf{89}, 213902 
%
\bibitem{ref42} A. Alu and N. Engheta 2005 \emph{Phys. Rev. E} \textbf{72}, 016623 
%
\bibitem{ref43} M. Silveirinha and N. Engheta 2006 \emph{Phys. Rev. Lett.} \textbf{97}, 157403 
%
\bibitem{ref44} A. Ciattoni, A. Marini and C. Rizza 2017 \emph{Phys. Rev. Lett.} \textbf{118}, 104301 
%
\bibitem{GH_ENZ} Y. Xu, C. T. Chan and H. Chen 2015 \emph{Sci. Rep.} \textbf{5}, 8681 
%
\bibitem{DrudeENZ} H. Wang, K. Du, C. Jiang, Z. Yang, L. Ren, W. Zhang, S. J. Chua, and T. Mei, {\it Phys. Rev. Applied} {\bf 11}, 064062 (2019).

\bibitem{Ale2013} A. Ciattoni, A. Marini, C. Rizza, M. Scalora, and F. Biancalana, {\it Phys. Rev. A} {\bf 87}, 053853 (2013).

\bibitem{Stockmann} M. H. Javani and M. I. Stockman, {\it Phys. Rev. Lett.} {\bf 117}, 107404 (2016).

\bibitem{Ale2017} A. Ciattoni, A. Marini, and C. Rizza, {\it Phys. Rev. Lett.} {\bf 118}, 104301 (2017).

\bibitem{Howard} A. Anopchenko, L. Tao, C. Arndt, and H. W. H. Lee, {\it ACS Photonics} {\bf 57}, 2631-2637 (2018).

\end{thebibliography}
\end{document}